\documentclass[11pt]{article}
\usepackage{geometry}                
\usepackage{graphicx}
\usepackage{ragged2e}
\usepackage{amssymb}
\usepackage{xr}
\usepackage{epstopdf}
\usepackage{amsmath, amsthm, amssymb}
\usepackage{amsfonts}
\usepackage{amsmath}
\usepackage{amsthm}
\usepackage{caption}
\usepackage{subfig}
\usepackage{float}
\usepackage{dcolumn}
\usepackage[dvipsnames]{xcolor}
\usepackage{url}
\usepackage{fancyheadings}
 \numberwithin{equation}{section}
\usepackage[sc]{mathpazo}
\usepackage{multirow}
\usepackage{setspace}
\usepackage{natbib}
\usepackage{appendix}
\usepackage{dcolumn}
\usepackage{color,hyperref,xcolor}
\hypersetup{
    colorlinks,
    citecolor=Violet,
    linkcolor=blue,
    filecolor=magenta,      
    urlcolor=NavyBlue,
}
\usepackage[utf8]{inputenc}
\usepackage{url} 
\usepackage{float}
\usepackage{subfig}

\usepackage{xr}
\usepackage{ragged2e}
\usepackage[T1]{fontenc}
\usepackage{lipsum}

\begin{document}	
\singlespacing

\title{{\Large \textit{WhatsApp Explorer}:\\ 
A Data Donation Tool To Facilitate Research on WhatsApp.}\footnote{Acknowledgements: the authors warmly thank developers Shreyash Jain, Shlok Pandey, Ayushman Panda and Ankit Kumar for their help in making WhatsAppExplorer a reality, Ved Prakash Sharma of \textit{Across India} and Mauricio Garcia of \textit{Connectar} for fielding multiple pilots (in India and Brazil, respectively) and for helping us identify workable strategies, as well as Jose Furones, the UC3M's data protection officer, and Laurence Tavernier, the ERC ethics scientific officer assigned to the POLARCHATS project, for multiple constructive dialogues that led us to identify the protocol we describe here. The authors acknowledge funding and/or logistical assistance from the European Research Council (Through the ERC POLARCHATS project), Knight Foundation, National Science Foundation, The University Carlos 3 of Madrid and Rutgers University. } \footnote{Authorship note: Both authors contributed equally to this research.}
\textnormal{\small}}

\author{
  Kiran Garimella\footnote{Assistant Professor, Rutgers University. Corresponding author: kg766@comminfo.rutgers.edu}
  \and
  Simon Chauchard\footnote{Associate Professor, University Carlos III Madrid \& Instituto Carlos 3 Juan March (IC3JM)}}
  
\date{March 27, 2024}

\maketitle

\centering
\includegraphics[width=.45\linewidth]{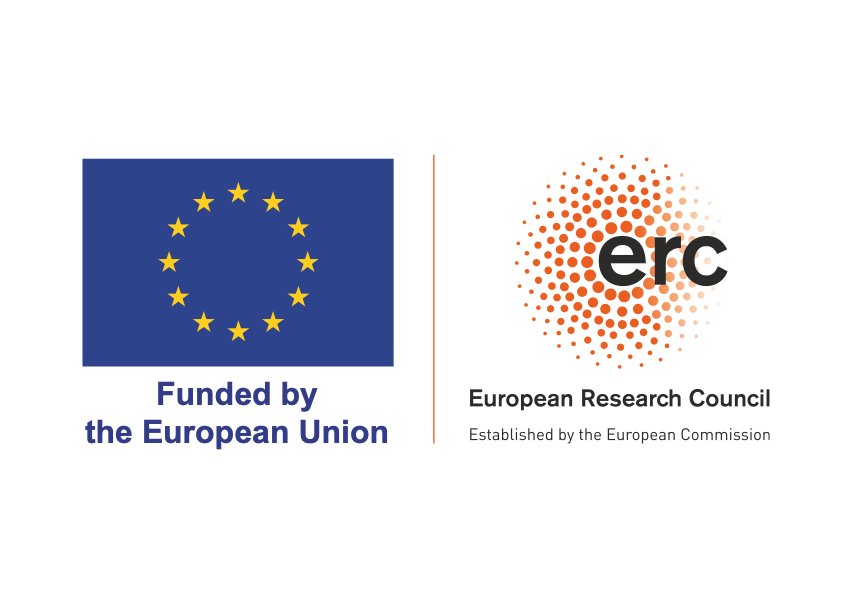}

\newpage
\begin{abstract} 
\normalsize
In recent years, reports and anecdotal evidence pointing at the role of WhatsApp in a variety of events, ranging from elections to collective violence, have emerged \citep{arun2019whatsapp}. While academic research should examine the validity of these claims, obtaining WhatsApp data for research is notably challenging, contrasting with the relative abundance of data from platforms like Facebook and Twitter, where user ``information diets" have been extensively studied~\citep{barbera2015tweeting,guess2019less}. This lack of data is particularly problematic since misinformation and hate speech are major concerns in the set of Global South countries in which WhatsApp dominates the market for messaging \citep{tucker2018social}. To help make research on these questions, and more generally research on WhatsApp, possible, this paper introduces \textit{WhatsApp Explorer}, a tool designed to enable WhatsApp data collection on a large scale. We discuss protocols for data collection, including potential sampling approaches, and explain why our tool (and adjoining protocol) arguably allow researchers to collect WhatsApp data in an ethical and legal manner, at scale. 


\end{abstract}

\pagebreak

\setlength\parindent{35pt}

\doublespacing
\justifying
\section{Introduction: The Need for WhatsApp Data}

Over the past few years, alarming media reports have pointed at the role of WhatsApp in a series of dramatic events. In countries such as Brazil and India, studies have indicated that group interactions on WhatsApp can alter the electorate's perceptions \citep{perrigo2019volunteers,bengani2019india,tardaguila2018fake}. Furthermore,  interactions on the platform have also been linked to even more dramatic behaviors, including, but not limited to, an increased likelihood of engaging in aggressive, extremist, or violent actions~\citep{chopra2019india,magenta2018whatsapp,ozawa2023disinformation}.


Although scholarship has started exploring these striking narratives, there is still substantial work to be done in assessing their accuracy and in unraveling the potential role of WhatsApp in these outcomes. Social scientists focusing on issues like hate speech or misinformation, particularly in the Global South, will increasingly require access to WhatsApp data, possibly on a large scale. To fully comprehend the platform's role in spreading problematic content and its repercussions, researchers need a deeper understanding of several aspects: (1) the nature and style of hate speech and/or misinformation shared on WhatsApp, (2) the volume of such content, (3) its virality, (4) the networks most conducive to its dissemination, and (5) the political, social, and contextual factors that give rise to this content and its tangible impacts. Moreover, it is essential for them to access this data in a manner that is practical, legal, and respectful of users' privacy.


Accessing WhatsApp data for research purposes however remains a challenging task. While evidence about the ``information diets" of Facebook and Twitter users has been available for some time now~\citep{barbera2015tweeting,guess2019less}, systematic evidence of a similar nature is notably lacking for WhatsApp, even though the platform's likely role in spreading problematic content if often acknowledged, particularly in the Global South~\citep{tucker2018social}. Furthermore, in the scarce instances where researchers do manage to obtain WhatsApp data, it often involves datasets too limited in scope to credibly address the previously mentioned questions, and/or the data is acquired in ways that might not be up to the strongest ethical standards.


How can researchers collect Whatsapp data from sufficiently diverse samples, and in ways that comply with legal and ethical requirements? Taking on this challenge, we introduce in what follows a donation tool and an associated protocol for harvesting substantial volumes of WhatsApp data. Our proposed data donation approach significantly reduces the practical difficulties associated with WhatsApp data collection, while adhering to prevailing privacy and ethical standards. We elaborate on our overarching strategy and outline a research protocol in Section~\ref{sec:protocol}. Section~\ref{sec:ethics} is dedicated to discussing the ethical considerations integrated into our approach. Subsequently, in Section~\ref{sec:sampling}, we discuss sampling. To lay the groundwork for these discussions, the following section begins with an examination of the challenges related to WhatsApp data collection. 


\section{Challenges of WhatsApp Data Collection}
\label{sec:challenges}

Researchers eager to engage in WhatsApp Data collection may face technical, legal, privacy-related, and practical challenges. 
Among these, certain challenges appear more daunting than others. Technical challenges are, for one, relatively limited: extracting data from private WhatsApp threads is technically easy once a thread participant (whether or not they are the admin of a group) consents to extract it: contrary to other platforms, WhatsApp makes it very easy for its users to archive the content of the conversations they are part of. Legal challenges originating from the platform are so far not a real issue. Of course, the platform's response to widespread research of the kind discussed in this article remains to be fully observed. However, there's a reasonable expectation that WhatsApp might view research aimed at identifying problematic behaviors or understanding their causes favorably. This likelihood is enhanced by WhatsApp’s own emphasis on encryption, which might limit the company's ability to scrutinize content on its own platform. We anticipate that external research, if it's carefully scoped and doesn't significantly compromise users' overall perception of privacy in WhatsApp conversations, may not only be legally permissible but also supported and encouraged by the platform.


That said, collecting WhatsApp data on a large scale still presents significant ethical, legal, and practical challenges that do require a robust discussion. These challenges are explored in greater detail in the subsequent subsections.


\subsection{Ethical and Legal Challenges}

Given that users can effortlessly export data from their threads, and considering researchers' inability to access private WhatsApp threads without the involvement of a thread participant, any WhatsApp data collection endeavor must essentially be a \textit{donation} effort. This requires one or more users to agree to share data from the threads included in their account and to undertake a series of steps to facilitate this export.


Donating data from one’s own account may be problematic from a privacy-protection point of view insofar as it may contradict guidelines, norms or laws protecting individuals’ privacy or limiting the processing of individuals’ personal data. The European Union’s General Data Protection’s Regulation (GDPR) currently constitutes the main example of such regulation, though privacy laws around the world – such as India’s Personal Data Protection Bill, Brazil’s LGDP or Canada’s Personal Information Protection and Electronic Documents Act (``PIPEDA") - echo most of the principles at the heart of the ruling when they exist. Besides, when it comes to the handling of users’ personal data, principles of user consent, anonymization and limitations on the use that can be made of such data will likely deserve a discussion, whether a local equivalent to the GDPR exists or not. 

In what ways might donating data from one’s own WhatsApp account exactly violate the privacy of users, as defined in these norms? Though we acknowledge that norms and regulations will differ depending on the case chosen by researchers, we believe that five potential challenges may be associated to a WhatsApp data donation program reliant on donations by what we hereafter refer to as a consenting “gateway user”: 

\begin{enumerate}

\item \textbf{Consent issues regarding third-party group participants}: these are users who aren't the gateway into a group, but whose data -- including phone numbers, profile pictures, messages, etc. -- appears in the threads. While the gateway user consents to share their data, they cannot provide consent on behalf of other users included in the dataset via their donation. This suggests that researchers should either obtain consent from these third-party participants, which might be impractical or methodologically undesirable, or ensure credible anonymization of the third-party data to render personal information unidentifiable.


\item \textbf{The extent of data collection allowed}: under the GDPR, this issue relates to the ``data minimization" principle, which dictates that personal data collection should be confined to what is directly relevant and necessary in terms of time and scope. Although the GDPR's Article 6 provides some exceptions for research deemed ``in the public interest," the minimization principle still limits the volume of data that researchers can and should gather, including if their research arguably qualifies as research in the public interest. In the context of WhatsApp, this may imply that data collection be limited to a relatively limited time frame. It may also indirectly raise the question of the type of threads (one-on-one vs. group, private vs. public, etc.) that may be harvested for research.


\item \textbf{The strategy for anonymization}: this refers to the methods employed to effectively anonymize stored data and minimize the risk of re-identification of both gateway and third-party users. Anonymization is crucial for preserving participant privacy, a priority in itself. However, it becomes particularly vital if the data is later shared with others, as is often the case under open science agreements. Ensuring anonymity is not only a privacy concern but also a requirement for broader data sharing in the academic community.


\item \textbf{Data protocols pre-anonymization}: this concerns the strategies that researchers will use for transferring and storing data \textit{before it is anonymized}. Specifically, the challenge lies in ensuring that this data, in its identifiable form, is not accessible to unauthorized individuals or lost prior to anonymization. Effective data management practices must be in place to secure the data during this vulnerable phase, safeguarding against both external breaches and internal mishandling.


\item \textbf{Managing ``unexpected findings" and legal disclosure obligations}: This pertains to how researchers will handle instances where they encounter data that may be subject to legal disclosure under international or local law. It's crucial to have protocols outlining the course of action for situations involving unexpected findings, especially those that might legally necessitate disclosure. This challenge involves balancing ethical research practices with compliance to legal requirements, ensuring responsible handling of sensitive or legally significant information discovered during the research process.


\end{enumerate}

\subsection{Practical challenges}

In addition to these ethical and, in some instances, legal challenges, there are also several \textit{practical} challenges that researchers must address to ensure the effectiveness -- from a research perspective -- of a WhatsApp data donation program. These challenges need to be navigated in accordance with the balancing principle described by~\citet{ohme2022digital}:


\begin{enumerate}
\item  The success of such a data donation program hinges on the \textbf{ease and efficiency of the donation process for potential donors}. If the procedure is overly tedious, costly, or time-consuming for the donor, it's more likely to fail. Additionally, the likelihood of failure increases if the donation protocol necessitates that an enumerator or research team associate has even superficial access to the data before it undergoes credible anonymization. Ensuring a smooth and private data transfer process is critical for both donor participation and data integrity.


\item The second challenge is \textbf{sampling}. To enable interesting research, a data donation program will in most cases need to convince a broad and ideally representative group of gateway users to donate their data. This could entail significant efforts and resources from the research team to ensure that the sample of donors is adequately diverse and representative, which is essential for the validity and reliability of the research findings.


\end{enumerate}

\section{Procedure and Tool: Introducing \textit{WhatsApp Explorer}}
\label{sec:protocol}

As noted above, to address these numerous challenges, the only available strategy is data \textit{donation}. In light of the outlined challenges, the strategy adopted must enable easy and straightforward donation. It should also safeguard privacy, foster trust among a varied pool of donors, and avoid legal complications for the research team.


We devoted a considerable effort to develop an effective solution to collect data with the aforementioned requirements from private WhatsApp groups.
This has led to the creation of a specialized web interface, \textit{WhatsApp Explorer}. \textit{WhatsApp Explorer} is designed to substantially mitigate the challenges previously discussed, while enabling the collection of large volumes of data that are valuable for research purposes.


\subsection{General Principles}

We first detail the broad principles of this strategy, before getting into the technical aspects of the tool, as well as the details of the data protocol, in the next sections. 

The core approach involves reaching out to individuals, encouraging them to contribute a portion of their WhatsApp data for social science research. Our described protocol centers on direct, face-to-face interactions between a ``gateway user" and a research associate. However, it's crucial to recognize the potential for adapting this to an entirely online process, enabling users to donate data remotely, without any in-person interaction. This focus on a face-to-face donation protocol is primarily because it seems more effective in our target regions for WhatsApp data collection, namely India and Brazil.\footnote{We focus on face-to-face interactions for two reasons: (i) our protocol requires gateway users scanning a QR code within their WhatsApp, necessitating a second screen device for autonomous data donation. This requirement might considerably limit our sampling pool, given the unlikelihood of most users owning two screens. (ii) Based on initial tests (elaborated on later), in-person interactions appear to foster trust between the research team and participants, potentially leading to greater and more varied user participation in the study.}


The primary technical innovation of our tool lies in its ability to streamline the data donation process for consenting participants. It allows users to effortlessly donate their data with the help of a research associate who facilitates the donation, without accessing the data themselves. To compensate participants for their time and contribution, we offer monetary incentives. Additionally, we assure extensive privacy and anonymization, emphasizing that their data, anonymized at the source, will not be shared outside the primary research team. Notably, in our proposed design, field staff such as enumerators and local partners will not have access to the collected data. The data collection is facilitated by these field staff but is directly encrypted and uploaded to a secure server accessible only to the Principal Investigators (PIs) and the core research team.\footnote{This approach also safeguards the research associates, as possession of certain types of WhatsApp content on their personal or professional devices could lead to legal risks if the data were to transit through their devices.}


Crucially, to mitigate privacy concerns, our approach avoids soliciting data from one-on-one conversations and small group threads. Instead, we focus on collecting data from larger groups. This decision is part of our strategy to balance the need for comprehensive data with respect for individual privacy, with more specifics outlined in the complete protocol provided later.


To address privacy concerns, we implement a thorough strategy for data anonymization \textit{during} the upload to our servers. We ensure that no raw, de-anonymized data is stored. For text data, we immediately anonymize any personally identifiable information, such as names, phone numbers, and email addresses, from the dataset. This anonymization is executed using advanced privacy-preserving algorithms from a reputable and widely used source: the Google Data Loss Prevention API.\footnote{https://cloud.google.com/security/products/dlp} This state-of-the-art tool is instrumental in maintaining the confidentiality of the data while enabling meaningful research analysis.


For visual content, our approach involves an irreversible anonymization process for most images and videos as they are uploaded. The exception to this rule is for content shared across at least k groups/threads (with k being a predetermined number, for example, 5). This practice ensures that we access only a minimal amount of un-anonymized visual content. Notably, the viral content that we retain and analyze is highly unlikely to be personal or private, as it, by definition, circulates widely across various online communities. The anonymization of visual content involves a multi-step process: (i) Automated Anonymization: we use automated tools to systematically blur faces and other identifiable features in images and videos, such as license plates; (ii) Human-Supervised Anonymization: following the automated process, we implement a second layer of anonymization, which is supervised by humans. This step occurs before any analysis of the data, reinforcing our extensive approach to anonymization and privacy protection.


\subsection{Technical Description}

Our tool leverages the capabilities of \texttt{whatsapp-web.js},\footnote{\url{https://archive.is/45ZNX}} an open-source library functioning as a WhatsApp client library for NodeJS. NodeJS, a JavaScript runtime, enables server-side execution of JavaScript, facilitating interaction with WhatsApp through its web browser application. The whatsapp-web.js library allows us to authenticate, read, and process messages on WhatsApp programmatically, functioning by operating WhatsApp Web in the background and automating its interaction. This automation relies on reverse-engineered API calls similar to those used by the official WhatsApp in its web application, enhancing the tool's robustness. The library whatsapp-web.js has a history of active development and contributions spanning nearly seven years. One of the drawbacks of our approach is the inherent dependence on the WhatsApp Web interface. Despite this dependence, our experience over the past two years indicates stability, as we have not encountered significant changes in the web interface that could disrupt our data collection process.


Our primary technical contribution is enhancing the existing library (whatsapp-web.js) to create a privacy-conscious, user-friendly front end, ensuring scalable and reliable data downloading. 
We start by implementing the automatic generation of a QR code, with the capability to regenerate the code upon expiration. 
Upon successful authentication using their WhatsApp account, users' authentication tokens are stored, offering permanent access unless manually revoked by the user. Utilizing these tokens, users can view a list of their group conversations, although our focus is on larger groups, this is not a limitation of the library itself. 
To accommodate multiple users and surveyors, we established a scheduled data backup system, enabling parallel downloads for scalability. This process, running nightly, automatically downloads new data for all users, encompassing all messages and various media types such as images, videos, documents, and audio files. This comprehensive approach ensures efficient and secure data collection on a large scale. As noted above, the data is first run through our anonymization procedures before it is stored.


The code can be accessed at 
\url{https://github.com/gvrkiran/WhatsAppExplorer}.
The code contains detailed instructions to setup an instance and configure the front and backend.
We also provide code which takes the data collected from the explorer to visualize the data. We provide the details below.\footnote{(i) In the current setup, unfortunately, the code still requires someone who is proficient in managing a Linux server and familiar with some programming.  (ii) To prevent potential misuse of our tools, currently, we limit the access only for academic and non profit use. The access can be requested by reaching out to the authors.}

To setup an instance of WhatsApp Explorer, researchers will need the following storage capacity: 1. A server with at least 8 GB RAM,16 CPU cores. Three ports need to be opened on the server for the frontend, backend and a monitoring dashboard.\footnote{We used ports 3000, 8000 and 8051 for the frontend, backend and the monitoring dashboards respectively.}
The code also includes a live monitoring dashboard which provides live statistics on how many users have been added and the amounts of data that was collected from them every day.
2.  Currently, we tested the tool on an Amazon AWS EC2 instance with 16GB RAM, 32 core CPU and 2 terabytes storage.
The instance does not use much RAM though to enable parallel data downloads, multiple CPU cores would help. For each active user being added to the system, WhatsApp Explorer consumes 1GB of RAM and 2 cores of CPU computation. This limits the number of concurrent users who can stay connected to the monitor at the same time, for instance, scaling up to large field deployment with dozens of enumerators might require a powerful server.



\subsection{Data Visualization Platform}
Once the data collection scripts are run every day (this is automated in our code above), we start another set of scripts to process the data to prepare it for visualization on dashboards. These dashboards visualize content which is spreading in the groups we monitor and help understand where what content was shared by whom, while at the same time not revealing any personal information of the users.
The pipelines perform various tasks, including clustering similar content, detect nudity, generating thumbnails from videos, etc.

We cluster the text, images and videos using different techniques to identify near similar versions of the same message. Text clustering is done using Locality Sensitive Hashing (LSH)~\cite{gionis1999similarity}, image and video clustering is done using Facebook's PDQ/TMK tools\footnote{\url{https://github.com/facebook/ThreatExchange}} respectively. These clustering techniques are robust to slight modifications and can find content which has been altered slightly (e.g. adding a water mark). 
Once the clusters are processed, we run through all the messages scraped from the data collection script stored in the MongoDB database. Here we extract links from text messages, and captions for media files (if any). We also generate persistent and unique identification codes for text messages, as media files are already assigned one during the data collection process. OCR is performed on images and video frames to extract text. Finally, we run a data indexer script to index the above data for elastic search.

Once the pipeline finishes its processing, we get the data hosted using an Express JS server and ElasticSearch server. This data is displayed using our dashboard (hosted at \url{http://analysis.whats-viral.me/}, available via login) which is developed over React JS, using the Material-UI library. This dashboard consists of various components such as user authentication, lazy data loading, dynamic search using filters, etc. The data which is displayed on the frontend is limited to the group the logged user is assigned to. Data is displayed in five individual tabs: Forwarded, Image, Video, Text, Link. Forwarded tab shows all messages (image, video and text) which were marked `Forwarded many times' by WhatsApp. The image, video, text and link tabs show content which was shared in at least 3 distinct groups in our dataset. To see the contextual information about any particular message, we can see all the groups where the message was sent and its respective timestamp. 
Data items are loaded in chunks as an infinity scroll to reduce internet consumption, and offer smooth user experience, instead of being overwhelmed by millions of data items.
All the communication with the backend happens using APIs secured with JWT Tokens. These tokens are also helpful for having persistent login sessions using cookies. Major backend operations such as user authentication, streaming video, loading metadata, etc. are performed by Express JS server, whereas fetch for lazy loading and search operations are handled by the ElasticSearch server.

\begin{figure}[h]
    \centering
        \includegraphics[width=\linewidth]{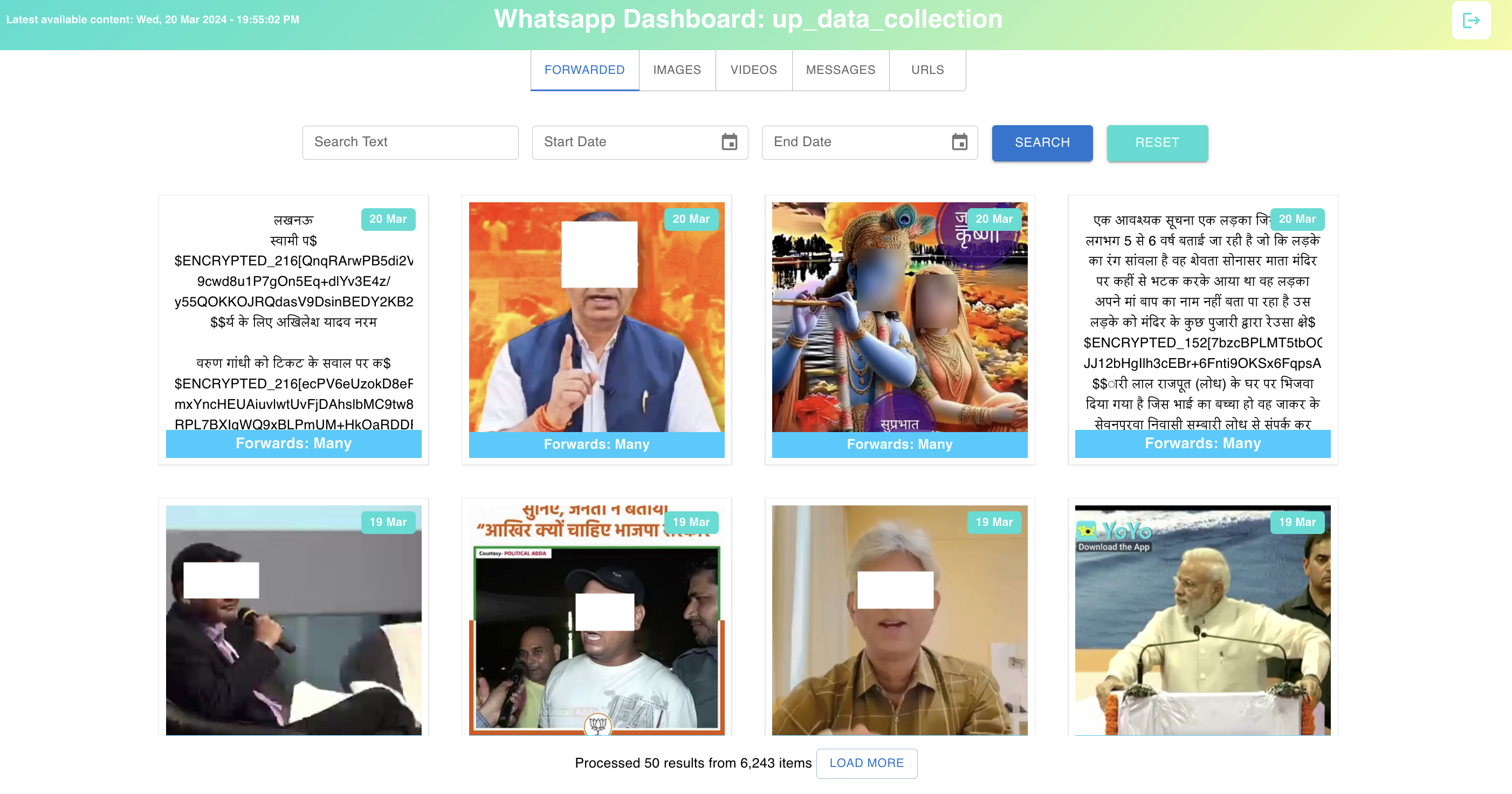} 
        \caption{A screenshot of the ``Fowarded" tab in our visualization dashboard. This tab shows text, video, and images which were `forwarded many times' on WhatsApp. We can see functionality to filter messages by date and search for text in images/videos.}
        \label{fig:whatsapp-explorer-visualization1}
\end{figure}

\begin{figure}
    \centering
        \includegraphics[width=\linewidth]{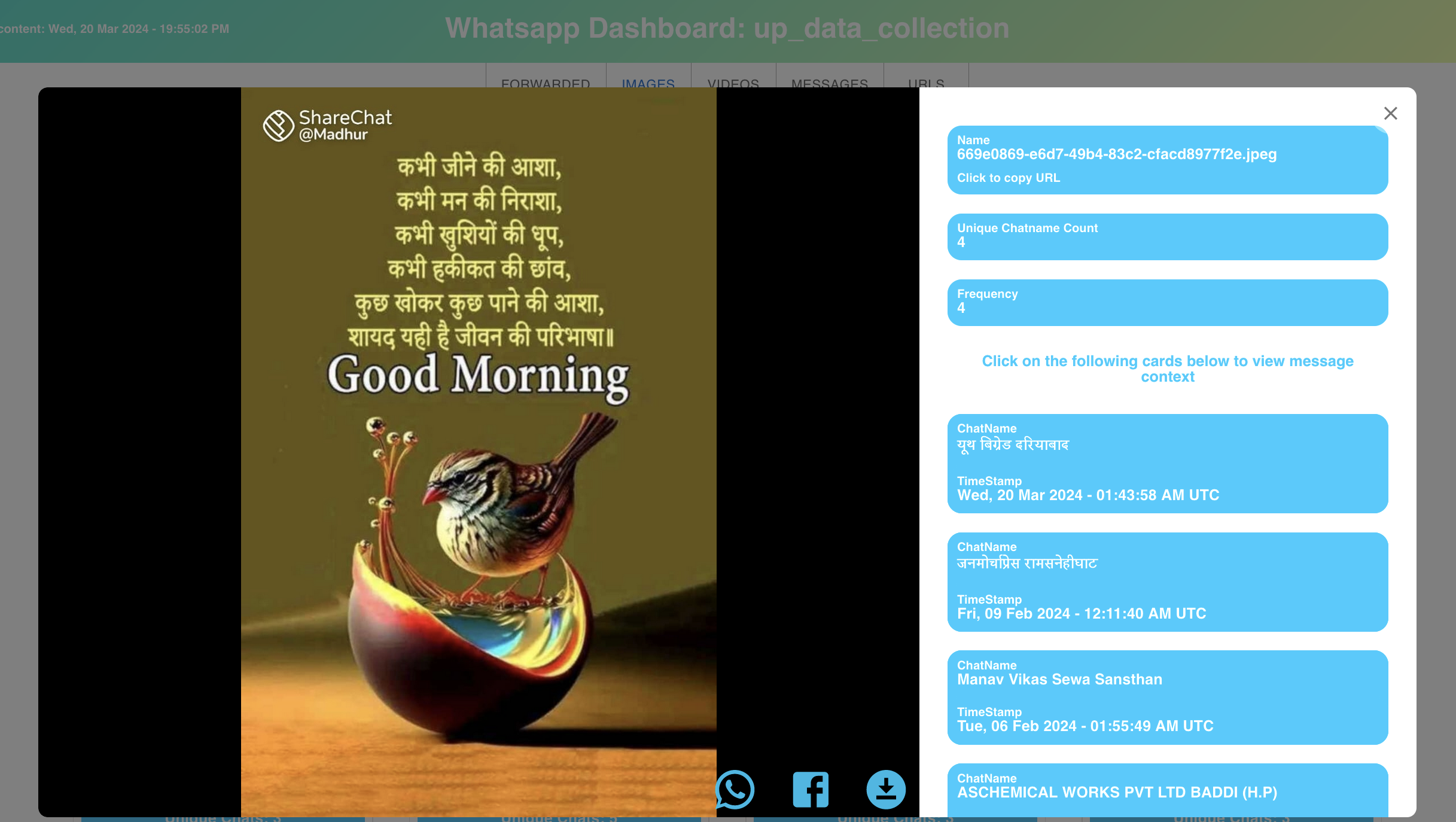} 
        \caption{Upon clicking a piece of content, this view shows the groups in which the content was shared along with their timestamps, giving a quick view of the spread of content.}
        \label{fig:whatsapp-explorer-visualization2}
\end{figure}

Figures~\ref{fig:whatsapp-explorer-visualization1},~\ref{fig:whatsapp-explorer-visualization2} shows screenshots of the data visualization platform.

Currently the authentication for the data collection and data visualization platforms is separate, to enable distinct stakeholders to have access to different views of the data, e.g., we could provide access to other researchers or fact checkers, while restricting access to other stakeholders like the surveyors.


\subsection{Illustration: a Protocol For Ethical WhatsApp Data Donation}

This tool may be used by researchers to assist with WhatsApp data donation. We stress however that using the tool does not guarantee that the data donation will be ethical and/or successful. Researchers will indeed need to design a specific protocol around it to meet these objectives. We detail here our own. 

As noted above, in the protocol we've developed for data collection in India and Brazil, we employ a strategy that involves the physical presence of a research associate. This approach is particularly suited for contexts in the Global South, where many users may not have the necessary hardware -- specifically, \textit{two} devices equipped with screens, such as phones, tablets, laptops, or desktop computers -- to independently complete the process online. Moreover, users might lack the technical skills required for an online process. Beyond the issue of hardware and skills, an in-person approach can be crucial in both the Global South and other regions for effectively conveying privacy assurances and building trust with participants.


While we cover potential sampling strategies at greater length below, this section concentrates on the procedure followed \textit{after identifying a potential donor}. Once a potential donor is pinpointed, a research associate, specifically a trained enumerator from a collaborating survey firm, personally visits the individual. During this visit, the associate invites the individual to partake in a research study focusing on their social media activity, and particularly with regards to discussion groups they are part of and about the content that circulates on these groups. 


This is how we envision the data collection will then look like, step-by-step:

\begin{enumerate}

\item 
The research associate, upon contacting an individual, explains the study’s objectives and seeks consent through a detailed consent process. This involves providing the potential donor with a flyer that contains essential information about the project. The flyer introduces the researchers and their objectives, including contact details. It features logos of all partnering organizations, both local and international. There is also a link to a registry detailing the research plans and legal basis, ensuring transparency in data processing activities. For further inquiries, a hotline phone number is provided. The flyer explains the anonymization strategy in clear, straightforward language, ensuring that it is understandable for non-technical readers. Additionally, it includes explicit instructions on how participants can opt out of the project at any stage.
    

\item If and only if the individual agrees to participate, the enumerator then asks them to scan a QR code generated by our tool through their WhatsApp app on their own smartphone. The research associate, using the web interface we've developed (found at \url{https://whatsapp.whats-viral.me/}), generates this QR code on their device. The participant can easily scan it using the ``linked device" function in WhatsApp, commonly used for connecting WhatsApp to a computer. It's important to note that at no point during this process does the research associate need to handle the respondent’s device, ensuring privacy and security.


\begin{figure}[H]
    \centering
    \includegraphics[width=.55\linewidth]{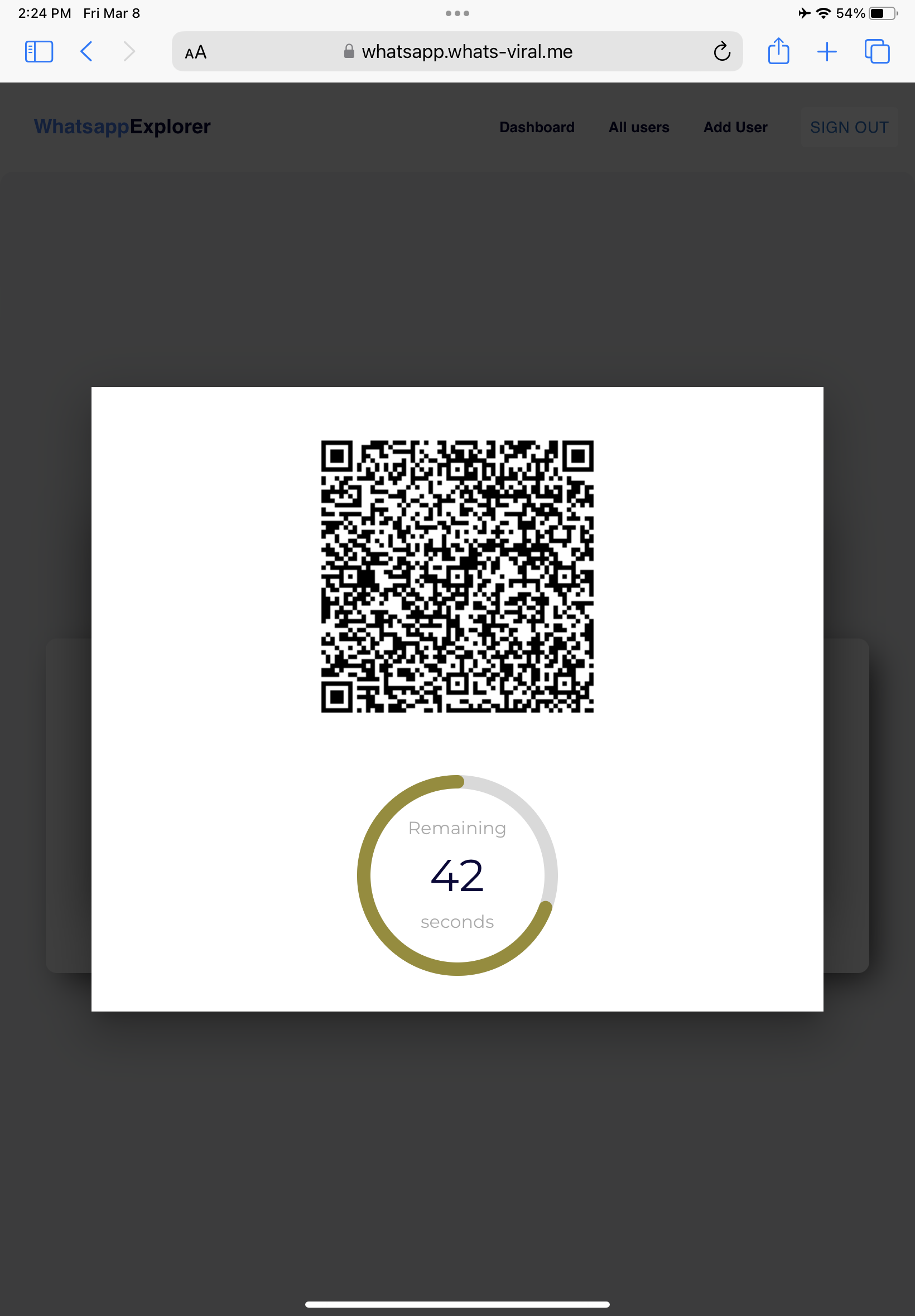}
    \caption{The scanning process}
\end{figure}

\item Once the QR code is scanned, the research associate connects the donor's WhatsApp account to \textit{WhatsApp Explorer}. During this process, the associate might briefly see the list of group included in the donor's account, but does not have access to any group content. Once the connection is established, the enumerator displays on their own device (likely a tablet) a list of these groups, then turns the device towards the donor for them to view.\footnote{At this stage, we automatically exclude smaller threads, defined as those with fewer than 4 participants, to minimize data handling and enhance privacy protection.
Throughout this entire process, enumerators do not have access to the content of the threads; the data are not uploaded onto their devices.} In the use we make of WhatsApp Explorer, screen shows as pre-selected all threads with 4 or more participants and that have at least 15 messages in the past two weeks, while all other groups remain deselected and unselectable.\footnote{These criteria are adjustable, depending on researchers' interpretation of the data minimization principle and any legal/ethical limitations they may face or choose to impose. The specified numbers are not definitive; they are perceived as a reasonable balance, but other configurations could also be appropriate, depending on their IRB's interpretation of legal and privacy guidelines.}


\begin{figure}[H]
    \centering
    \includegraphics[width=.6\linewidth]{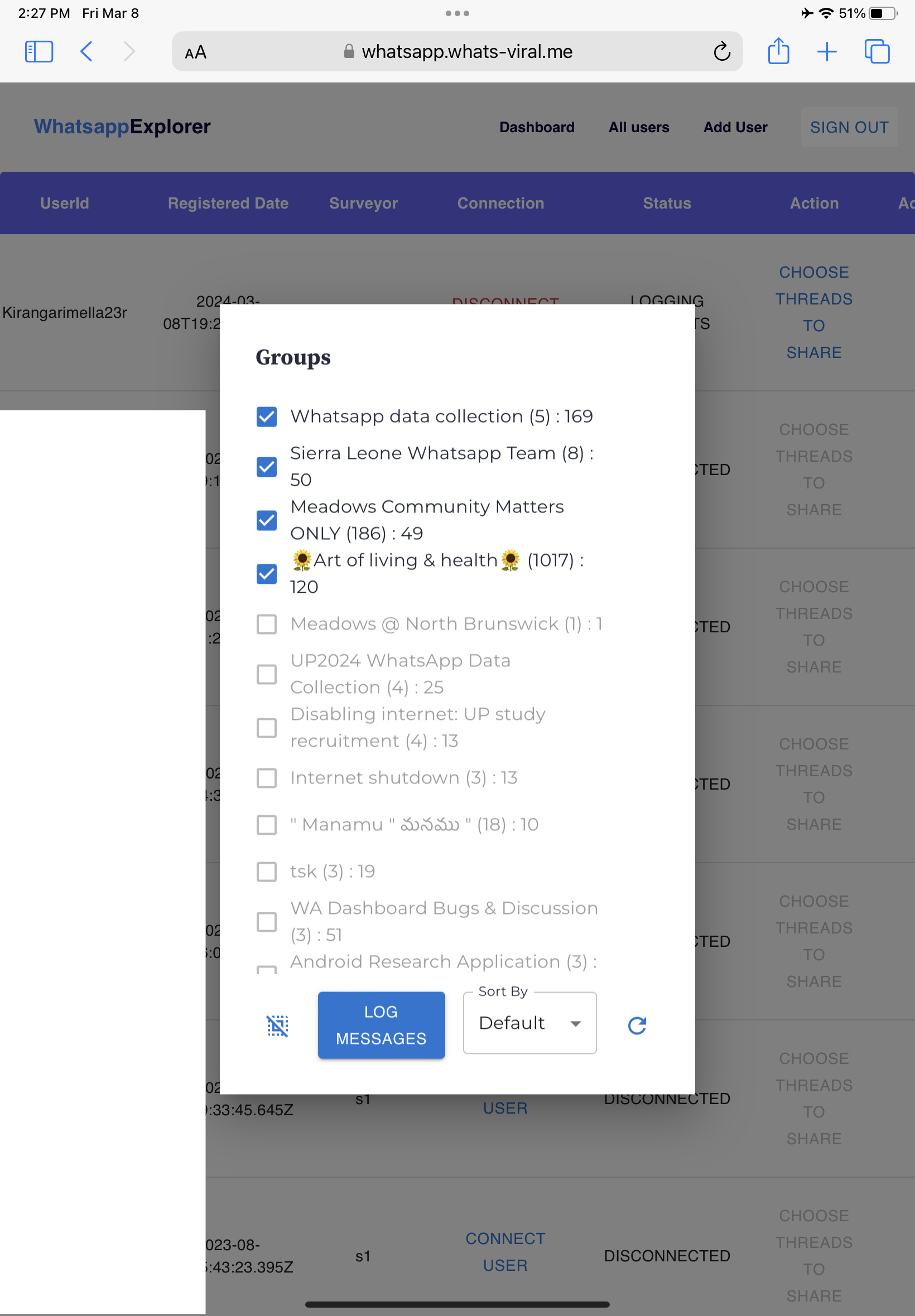}
    \caption{Including and excluding groups to donate}
    \label{fig:brazil_mean_consented_groups}
\end{figure}

\item At this stage, participants can actively de-select any number of the pre-selected groups they may not wish to donate, by clicking the respective boxes; alternatively, they can choose to donate all pre-selected groups. By doing so, they consent to share data from their chosen groups for a period extending two months prior and two months after their interview date. Participants thus have the flexibility to limit their donation to either historical or future data, or a combination of both, from selected groups. During this selection process, the interviewer inquires about the type of data the participants are comfortable sharing:
a) Historical data (from the past two months), b) Future data (for the coming two months), c) Both historical and future data. If participants opt to share future data, they are informed about the necessary steps to ensure data collection continues for the next two months and how to stop it, should they decide to do so. These actions can be swiftly executed on their own phones. Detailed instructions about this are provided in the draft flyer distributed during the consent process. The research associate is also prepared to demonstrate these steps in person if needed.


\item Once the participant completes the inclusion/exclusion process, the research associate presses the “log messages” button, triggering two critical actions. Firstly, the content from the last two months of the chosen threads is uploaded to our secure server, having been anonymized according to our established strategy. We strictly adhere to the policy of uploading and storing only anonymized content, which is encrypted during its transfer from the field staff’s devices to our cloud server. This step assures that neither co-PIs nor research staff have access to any non-anonymized content. Secondly, the action sets up an anonymized mirror copy of the selected threads, enabling the collection of data from these threads \textit{going forward}.
Ensuring a finite period for future data collection is paramount, hence the system is programmed to automatically deactivate after two months. This automatic shutdown leads to the user's disconnection and the deletion of their contact information from our database, effectively and securely concluding their data contribution.
 

\item At this stage, once the messages and threads have been logged, the respondents are asked to answer a brief series of questions. These questions pertain to up to 5 of the threads for which data were either harvested or excluded. Additionally, the respondents provide information about their own demographic characteristics. 


\begin{figure}[H]
    \centering
    \includegraphics[width=.7\linewidth]{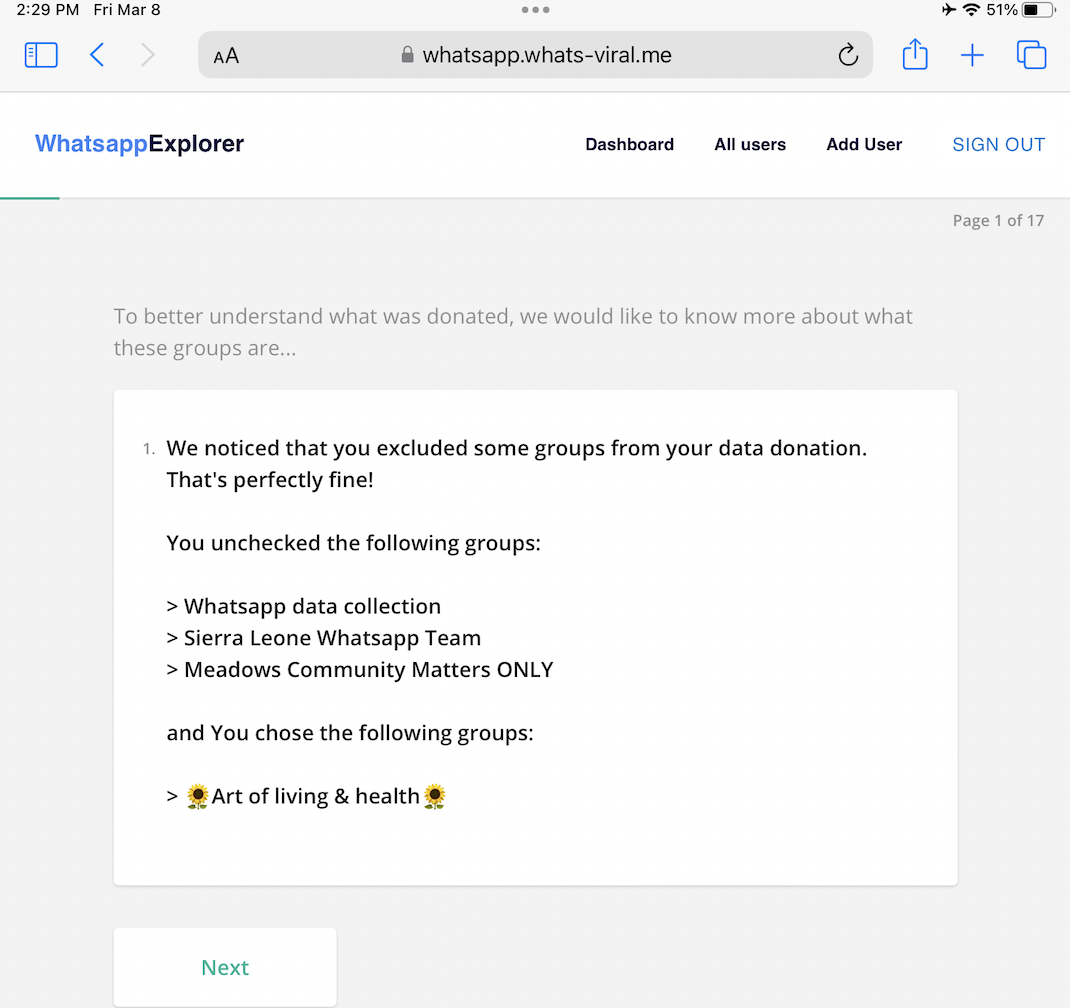}
    \caption{Screenshot from post-donation survey}
\end{figure}

\item Within a few weeks of their participation, the respondents are compensated for their contribution, typically receiving rewards such as phone credits directly on their phones. Along with this compensation, they are once again provided with the contact information for the “hotline” and a link to further details about the data donation program.


\end{enumerate}

\section{Legal and Ethical Safeguards}
\label{sec:ethics}

Our protocol includes clear legal and ethical safeguards, which we detail here.

\subsection{Strong Anonymization and Privacy Protection}

Understanding the effectiveness of our anonymization strategy in protecting user privacy requires a comprehensive explanation of the data collected, along with the specifics of how and when anonymization occurs. This detailed overview will provide clarity on the types of data harvested through this approach and the precise methods and timing employed to ensure privacy is maintained throughout the data collection and analysis process.


With this data collection process, we gather a range of information. This includes details on the users' connections (i.e., their address book), the identities of individuals they have communicated with on WhatsApp, the number of groups they are a member of, the size of these groups, and, most critically, the content of chats to which users have consented to share. The content from these chats encompasses messages, images, and videos, along with the timestamps of these communications and the identities of the senders.\footnote{Additionally, as will be detailed later, users' names and contact numbers are immediately anonymized in a manner that ensures neither we nor anyone else can access this personal information.} This approach is designed to balance the need for comprehensive data collection with strict adherence to privacy and anonymization standards.



Our data collection process involves multiple stages of anonymization to ensure privacy. The first stage, an automated anonymization process, occurs before the data is stored on our servers. During this stage, we employ automated procedures to strip any personally identifiable information, such as names, phone numbers, and emails, from the dataset. This is accomplished using advanced, privacy-preserving algorithms from a well-established and widely utilized library provided by Google, known as the Google Data Loss Prevention API.
Each piece of sensitive information is converted and replaced with a unique identifier. Although this identifier could technically be used for re-identification, we eliminate the original key as soon as we have confirmed that the data is securely stored, making re-identification impossible. Additionally, we do not store group icon pictures (i.e., profile pictures) in our dataset. 
These measures are integral to our commitment to maintaining the highest standards of data privacy and security throughout our research process.

 
In addition, for pictures/videos included within the threads that we collect, we create the following pipeline: 

\begin{enumerate}
\item At the outset, we securely store the pictures and videos collected on our servers. Simultaneously, we generate hashes for each of these visual elements. Hashing is a process that transforms these images and videos into a unique, fixed-size string of characters.
\item Following the creation of hashes for the images and videos, we utilize these hashes to determine whether the same image or video appears in the data shared by multiple users. Since each hash is a unique identifier corresponding to a specific image or video, comparing these hashes enables us to identify instances of the same content being shared across different data sets.
\item We take significant measures to ensure the privacy and anonymity of the visual content we store. Most images and videos are irreversibly anonymized, with the exception of those shared by at least 5 groups or threads in our dataset and that do not contain personal data. Given that this will apply to a relatively small subset of the content, our core research team will review each of these items individually to determine their eligibility for exclusion from anonymization. This thorough procedure is designed to ensure that we avoid accessing the vast majority of un-anonymized visual content, thereby maintaining strict adherence to privacy standards.
\end{enumerate}

For the anonymization of visual content, we employ state of the art tools designed to automatically blur faces. This technology offers a convenient and automated process for anonymizing not only faces but also other identifiable features in images and videos, such as car license plates. This entire procedure occurs within our servers, ensuring that the data never leaves our system during the anonymization process.
Once the images and videos are anonymized, we replace the original, un-anonymized files stored on our servers with these newly anonymized versions. Following this, the original un-anonymized images and videos are permanently deleted. This step is crucial to ensure that no personally identifiable visual content is retained in our database, aligning with our commitment to uphold strict privacy standards in our research.

 
Despite the rigor of our protocol, it's important to recognize that perfect, infallible anonymization is unattainable and cannot be solely reliant on automated procedures. Therefore, we introduce a second layer of anonymization, this time driven by human oversight. This Systematic Anonymization Audit (SAA) is conducted before any analysis of the data. The objective of this audit is to enhance and, ideally, perfect the anonymization process already in place. The SAA involves a meticulous review of the data by trained personnel, who manually check for any residual personally identifiable information that might have escaped the automated procedures. 


In practice, the SAA involves research associates, well-versed in the relevant context, examining all text and visual content previously anonymized through automated means, to assess any remaining re-identification risks of personal data. The focus initially is on the text content, where they systematically eliminate any references to specific locations, identification numbers, and mentions of individual attributes, whether physical, physiological, genetic, mental, economic, cultural, or social. Additionally, details about unique possessions or aspects of an individual's company or social network that could lead to identification are also carefully redacted. This process is crucial for enhancing the privacy safeguards already established by the automated anonymization methods.


After completing the text anonymization, the focus shifts to further anonymizing visual content as necessary. Specifically, this involves blurring any elements in the images that could hint at or directly indicate a location, such as street signs or storefronts. Additionally, we ensure that potentially distinctive or atypical landmarks in the background are blurred. If the attire of individuals in the images is distinctive or identifying, that too is blurred for added anonymity. Finally, any distinctive body marks, like scars or tattoos, are also blurred, although faces will have already been obscured in the earlier stage of anonymization.
We will update this list as need occurs.


\subsection{Policy with Regards to Legally Problematic Content}

Our protocol in addition incorporates explicit provisions to address scenarios where data collection or research activities might lead to the discovery of content that requires mandatory legal disclosure, in accordance with the laws of the countries where our research is conducted. 


Although we anticipate a low likelihood of such occurrences ex-ante, we acknowledge that this data collection project or the subsequent analysis could result in ``unexpected findings." These are findings that, while outside the primary scope of our research objectives, might require researchers to take specific actions, such as disclosing information to appropriate or designated authorities. This aspect of the protocol ensures we're prepared to responsibly manage any unforeseen or incidental discoveries that arise during the course of our research.


In a project centered on social media and violence, ethical dilemmas classified as 'serious and/or complex' might arise if the research uncovers unexpected or incidental findings that necessitate interventions for the safety and well-being of participants. This could include indications of physical abuse, self-harm, drug dependency, or neglect, particularly in minors. Additionally, the research might reveal information subject to mandatory disclosure under the national laws of the countries where the research is conducted. Such situations would require researchers to break the confidentiality normally afforded to research participants. Instances that might necessitate such disclosures include criminal activities like crimes, child sexual exploitation, human trafficking, or acts of terrorism.


Should such content be detected, our research team commits to consulting with the ERC POLARCHATS' ethics advisory board, which includes specialists from both Brazil and India, within 3 days from the time of discovery. This consultation will focus on determining the most appropriate course of action for each individual case. We consciously avoid setting a predetermined, blanket policy for such situations, given the dynamic political environments in both countries and the possibility of political bias within their judicial systems. This approach allows for a more nuanced and context-sensitive response to each unique scenario, ensuring ethical integrity and adaptability in our research practices.


\subsection{Restraint in Amount of Data Collected}

As previously mentioned, we request participants to share data spanning up to four months: two months prior and two months following their involvement. This data is specifically from threads meeting certain criteria--those with four or more participants and at least 15 messages in the past two weeks. However, participants are also provided the flexibility to share only a portion of this selected data. They have the option to contribute either historical data or data going forward. Additionally, they can choose to exclude any thread from the initial list presented to them and can opt out of as many threads as they wish. It's also important to note that participants have the freedom to withdraw from the program at any point after giving their consent, ensuring their ongoing control over their data and participation.


Consequently, our approach does not entail an indefinite connection to participants' WhatsApp accounts, and we deliberately refrain from collecting data from certain types of threads. We consider these parameters to strike an appropriate balance between three key elements: (i). adherence to the data minimization principle, (ii). the practicality of our extensive anonymization strategy, and (iii). our capacity to conduct meaningful scientific research, particularly statistical analyses, that serve the public interest. This balance is critical to ensure that our research methodology is both ethically sound and scientifically robust.


All other threads, particularly one-on-one conversations, are completely omitted from our data collection process. These threads do not appear on the research associate's screen when interacting with potential donors and cannot be selected for inclusion in the donation, even if the gateway user expresses a desire to share them.


These measures substantially restrict the volume of data that our research team can collect. Although this approach is designed to furnish a considerable amount of politically and socially pertinent data regarding users' WhatsApp activities in the target countries, enabling meaningful and representative analyses, it also serves as a safeguard against the excessive or unrestrained accumulation of private data. We view this restraint as essential, particularly as data donation programs become more prevalent.


The ongoing discussion within the research community about the optimal balance between data access for research purposes and privacy considerations is crucial. We believe that the provisions we have implemented significantly address the ethical and legal challenges associated with WhatsApp data collection, as outlined earlier. However, it's important to recognize that while our tool and protocol effectively navigate these challenges, they don't completely eliminate all obstacles to deploying a successful data donation program. Adhering to ethical data collection practices doesn't inherently assure, and might even conflict with, the ability to gather data from samples that are scientifically compelling and representative.


\section{Solving the Sampling Challenge}
\label{sec:sampling}

As noted above, our protocol imposes strict limitations on the types of data,  the types of groups, and the amount of data we collect at the individual level. We believe these limitations to strike the right compromise between our ability to carry research and the need for privacy protection, insofar as these limitations theoretically do not prevent us from collecting the data we are most interested in, in the context of our research; that is, the content of larger groups (here defined as groups including 5 or more members). 

With this said, even if we deem the data donated \textit{by each individual donor} to be interesting enough, our ability to understand what circulates on large WhatsApp groups in India and Brazil is contingent on our ability to solve the aforementioned sampling challenge. That is, it is contingent on our ability to gather data from samples of users that are ideally representative of the societies we are interested in (or at least of \textit{some} subgroups of users in these societies), or as a second best, samples which at least include donors from diverse enough subgroups from these populations. 

In the rest of this section, we outline several possible strategies that we deem more or less desirable and/or realistic, based on the pre-tests and exploratory surveys we have so far conducted in India and Brazil.\footnote{Because our main research objectives are openly descriptive, and require that we speak about a broader population, we exclude from the get-go snowball sampling as well as simpler forms of convenience sampling.}

\subsection{Probability Sampling}

A first -- and maybe ideal -- possibility may be to rely on probability or random sampling, possibly based on some stratification. In this case, after the research team identify targeted individuals based on some form of random selection from an established sampling frame (a population list), associates would approach them and ask them to donate part of their WhatsApp data, following the protocol detailed above. This attempt at extracting WhatsApp data may be embedded in another survey or constitute the central objective of the study. 

Although others may be more lucky in different locations, our experience suggest that such a strategy, while not altogether impossible, may be either expensive or inefficient or both. We attempted a version of this in August 2023 in a single district of Uttar Pradesh, India, initially over a large sample.\footnote{While the initial plan was to contact 800 individuals, we cut short the experience after 6 days, in light of low participation rates.} Despite the fact that we were offering relatively large incentives, the proportion of randomly selected respondents that were contacted and who later accepted to donate some of their WhatsApp data never crossed the 15\% mark. Importantly, while this was not encouraging, the denominator here included a large a amount of individuals (more than half of those contacted, in fact) who either (i) did not own or have access to a smartphone, (ii) had access to a smartphone but did not have internet data at the moment of contact, or, (iii) had a smartphone and data but either did not use WhatsApp or used it too little to qualify as eligible donors. We also note that the rate of donation was almost three times higher among some demographics that may be of specific interest to researchers (young men, for instance, in this context), potentially making random sampling a viable option if targeted on some specific demographic subgroups. Nonetheless, such numbers would require researchers to deploy very large resources in order to obtain donations from a sample of individuals that would  eventually remain a biased sample.
While we did not carry a similar experiment in Brazil, we embedded in two existing studies based on stratified random sampling (first, in the relatively affluent city of Blumenau (N=426), in the South of the country; and separately across the Northern state of Pernambuco (N=999)) a series of attitudinal questions gauging individuals' willingness to donate their data. Importantly, these questions were asked after the protocol was explained in great detail, and respondents were able to consult all formal documents that our potential donors would receive in the final study. They were also told that they would receive relatively large monetary incentives immediately before they were asked whether they would accept to participate to the data donation program. Finally, to make the question even more meaningful, respondents were asked if the research associate could return the following week at an agreed upon time to complete the data donation protocol. While the actual rate of donation would likely be smaller than what we can infer from these behavioral intentions, results were overall more encouraging than in India. In our first sample, 25\% of the individuals we overall approached pledge to donate their data, and 29\% of those who declared being active on some WhatsApp groups. Further, the number rose, sometimes as high as 40\% among some demographics, with the young, male and less affluent respondents overall being more prone to donating their data.\footnote{We by contrast detected no correlation with education.} While it took place among a less affluent population, with much lower rates of WhatsApp usage, results were comparable in our second sample: 18.3\% of the individuals we overall approached pledge to donate their data, but this number grew to 35\% among those who declared being active on some WhatsApp groups and were eligible donors as per our criteria. Importantly, differences in willingness to donate overall did not significantly differ across age, gender, education level or income levels in this second sample. 


Altogether, while more research may be needed to increase our confidence in this intuition, these data suggest that random sampling might be a feasible option if researchers can offer incentives as substantial as ours, if they are able to previously identify the criteria of likely donors, and if these correspond to a population of interest. This may be possible in some contexts and with regards to specific research questions. Nonetheless, we believe this should overall remain a difficult and expensive option, including for researchers with substantive research budgets. For this reason, we believe researchers may be better off using a less probabilistic and more creative approach to sampling.

\subsection{A More Feasible Alternative: Decentralized quota sampling}

Given the challenges we faced as we attempted to rely on random sampling, we devised an alternative approach for our pilot studies in India and Brazil. 

While it does not have the obvious advantages of random sampling, the strategy we relied on India avoids some of the most obvious pitfalls associated to purposeful sampling, convenience sampling or snowball sampling. We name this strategy ``decentralized quota sampling". Practically, we recruit a large number of research associates (a much larger number of associates than may be necessary in a traditional survey) and later request each of them to recruit, following some strict demographic quotas implemented at the level of each research associate, a relatively small number of donors located around their own place of residence. Concretely, we have many research associates disseminated around a territory of interest, each of which recruits a relatively small number of donors following some quotas. This strategy guarantees that we recruit individuals with sufficiently diverse demographic characteristics -- quotas prevent homogeneity in the pool of donors recruited by each associate --, and that we recruit individuals drawn from a variety of different networks, as each associate may only recruit a small numbers of donors. Provided the number of potential donors recruited by each associate remains small (for instance $<$ 20), and that the number of associates initially recruited is high and spread over a series of locations which may jointly be representative of the population of interest, we contend that such a strategy would enable researchers to collect data based on larger (and possibly more diverse) samples than random sampling arguably would. More importantly, such a strategy would allow researchers to better focus their resources. 

We piloted this strategy in one district of Uttar Pradesh (India) in September 2023. This allowed us to receive donations from 379 users giving us 1,094 groups at a fraction of the cost we would have spent had we relied on random sampling. We later piloted this strategy in Sao Paulo (Brazil) where we harvested data from 201 users, who gave us a total of 792 groups.

While we have have so far refrained from engaging in analyses about the content of the threads we exported, these pilots already provide us with important aggregate statistics about WhatsApp usage and about willingness to give away content in both of our samples. As shown in Figure 6, the average number of groups with more than 2 participants that donors had on their phones was 23.2 in Brazil and 15.85 in Brazil.\footnote{For the record, the average number of one-on-one conversations that donors had on WhatsApp was 171.42 in Brazil and 140.17 in India}. Of these, users were by design - following the limitations we self-impose as per our protocol - only able to donate an average of 5 and 3.32, respectively. One good news from a sampling standpoint is that they did give away the vast majority of these groups, as evidenced by the comparison between the number of eligible groups and the number of effectively donated groups. While our sample is likely biased, there is little evidence that the sample of groups that participants donated was.
Figure~\ref{fig:group_sizes} in turn shows the distribution of group sizes in our two datasets. The median group sizes in India and Brazil were 104 and 71 respectively. As we can see, these are potentially large groups as indicated by previous research on the membership of users in large public groups~\citep{lokniti2018}.
Because we collect messages over a period ranging from 2 to 4 months per participant, this strategy allows us to harvest a large number of messages: we harvest an average of 2,760 messages per donor in Brazil, and 1,103 in India. 

\begin{figure}[H]
    \centering
    \includegraphics[width=.85\linewidth]{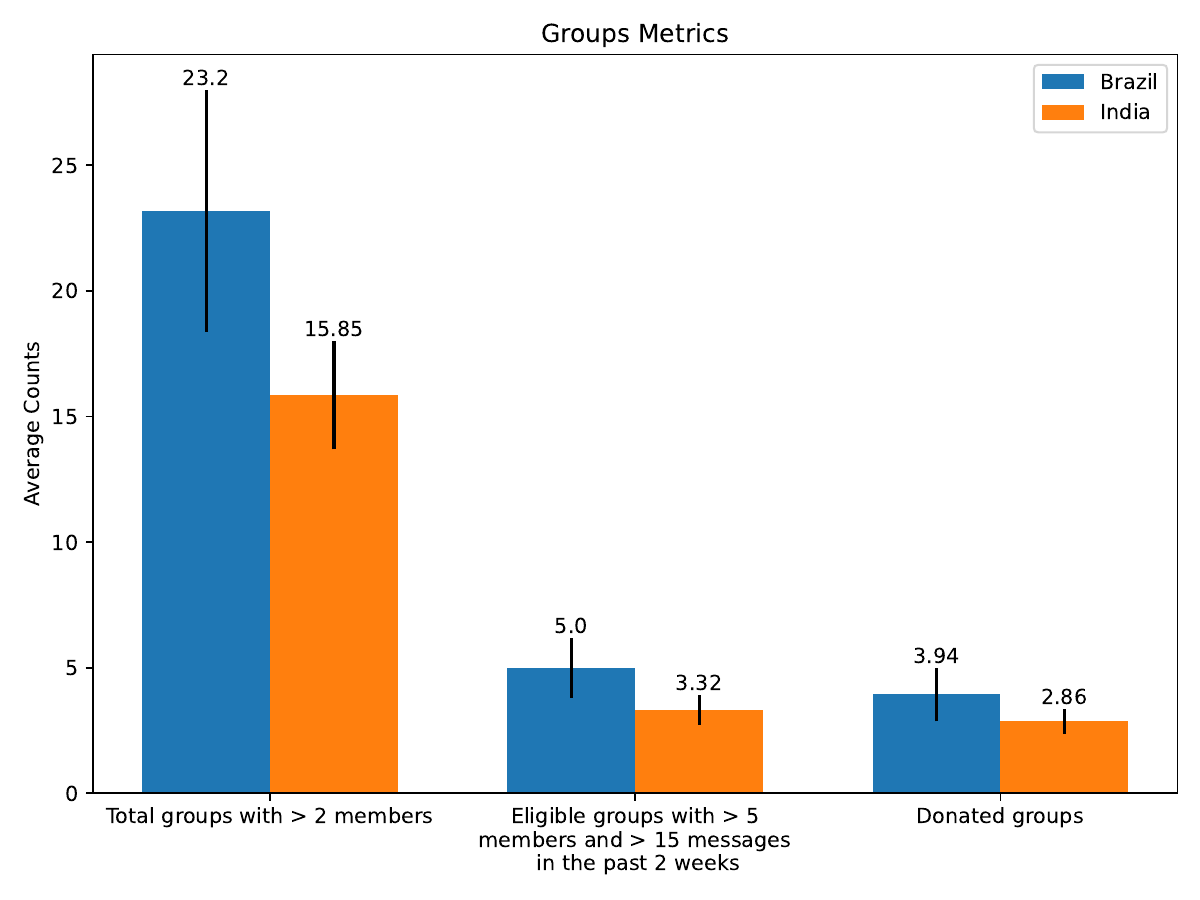}
    \caption{Group metrics}
    \label{fig:group_metrics}
\end{figure}

\begin{figure}[H]
    \centering
    \includegraphics[width=.9\linewidth]{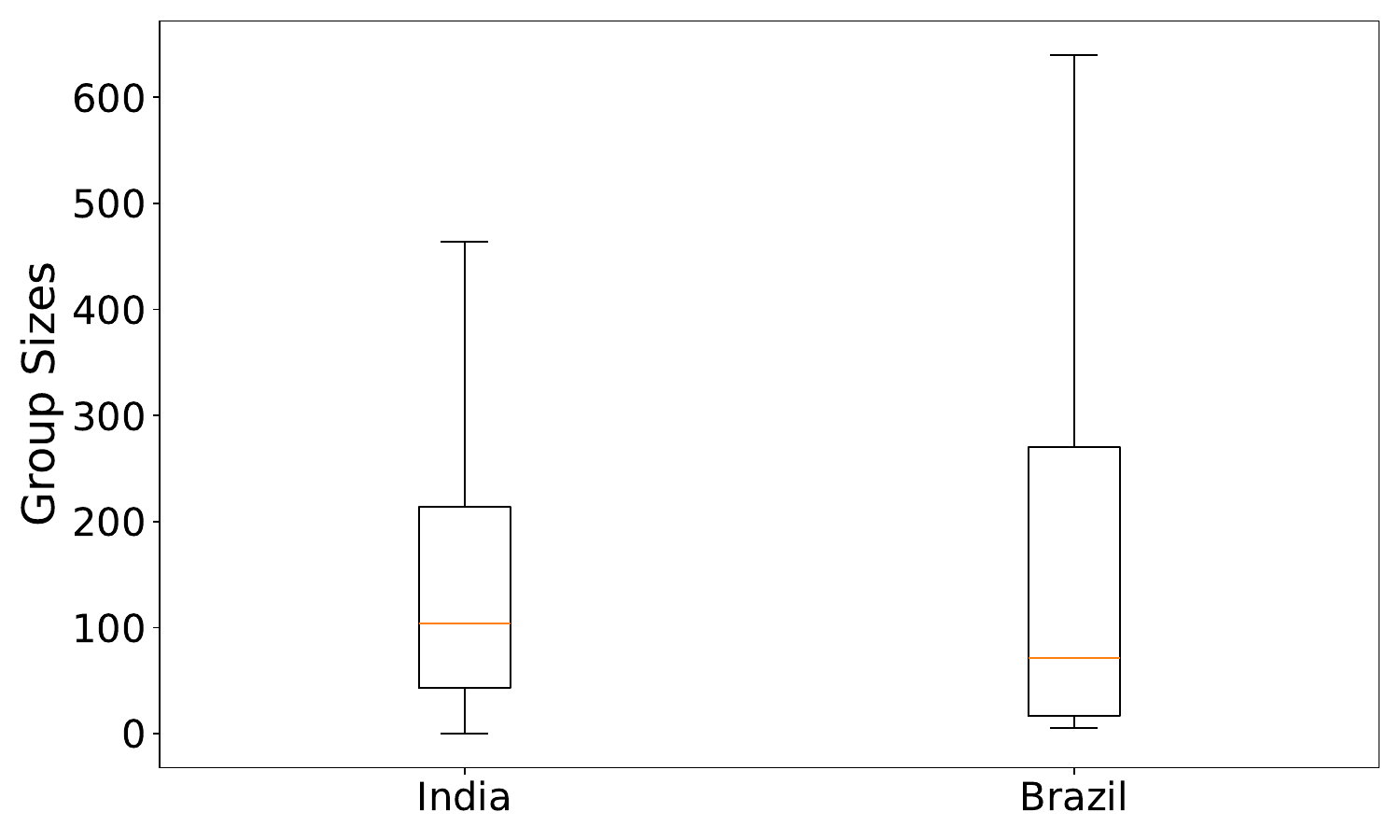}
    \caption{Size of Donated Groups in India and Brazil}
    \label{fig:group_sizes}
\end{figure}



Breaking this down by age category (Figure 8), we confirm that all age categories maintain relatively few large and active groups (as implied by Figure 6), but also show that the number of such groups broadly seems to correlate with age in both countries, with the youngest donors having and hence donating the lowest number of groups.\footnote{Small N in some of the highest age categories in both countries make these aggregate statistics potentially unreliable.} 

\begin{figure}[H]
    \centering
    \includegraphics[width=1\linewidth]{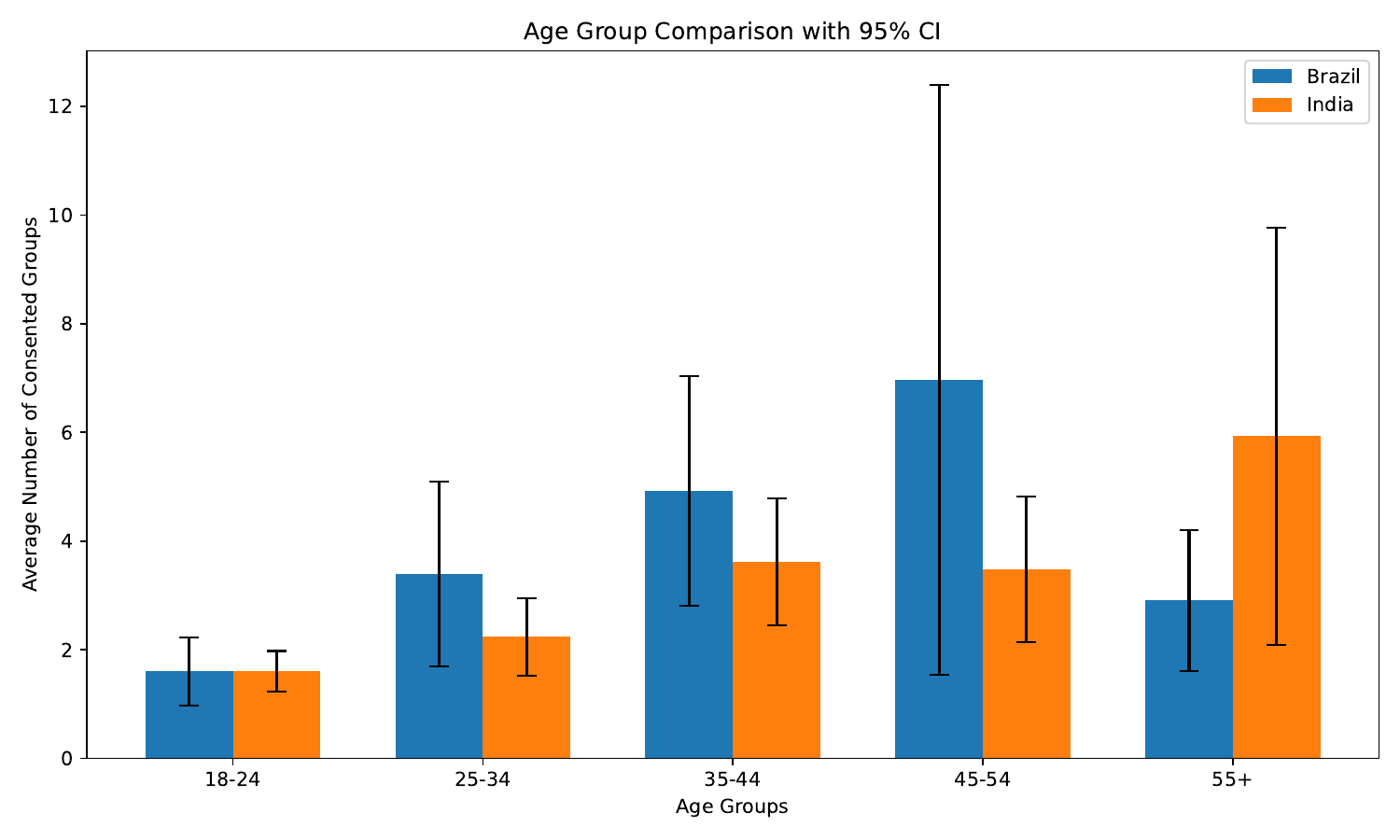}
    \caption{Number of donated groups by age categories}
    \label{fig:age_consented}
\end{figure}

The comparison Brazil/India in terms of activity on large WhatsApp groups so far may imply that this activity is somewhat correlated to living standards, with comparatively better-off Brazilians maintaining a higher number of such groups. A break down of the number of donated groups by demographics other than age (in Figures 9 and 10) broadly seems to confirm this point. In India (Figure 9), while we did not have a reliable way to measure donors' income as part of the survey tagged onto our pilot, we see that the number of groups is extremely correlated with caste categories, with upper-caste donors maintaining many more large active groups than lower-caste donors. Similarly, Hindu respondents appear to maintain (and hence donate) many more groups than Muslim respondents. In Brazil (Figure 10), we observe similar evidence suggesting that income and education correlate to the amount of groups. Meanwhile, we detect no clear correlation with gender, race, or religion. 

\begin{figure}[H]
    \centering
    \includegraphics[width=1\linewidth]{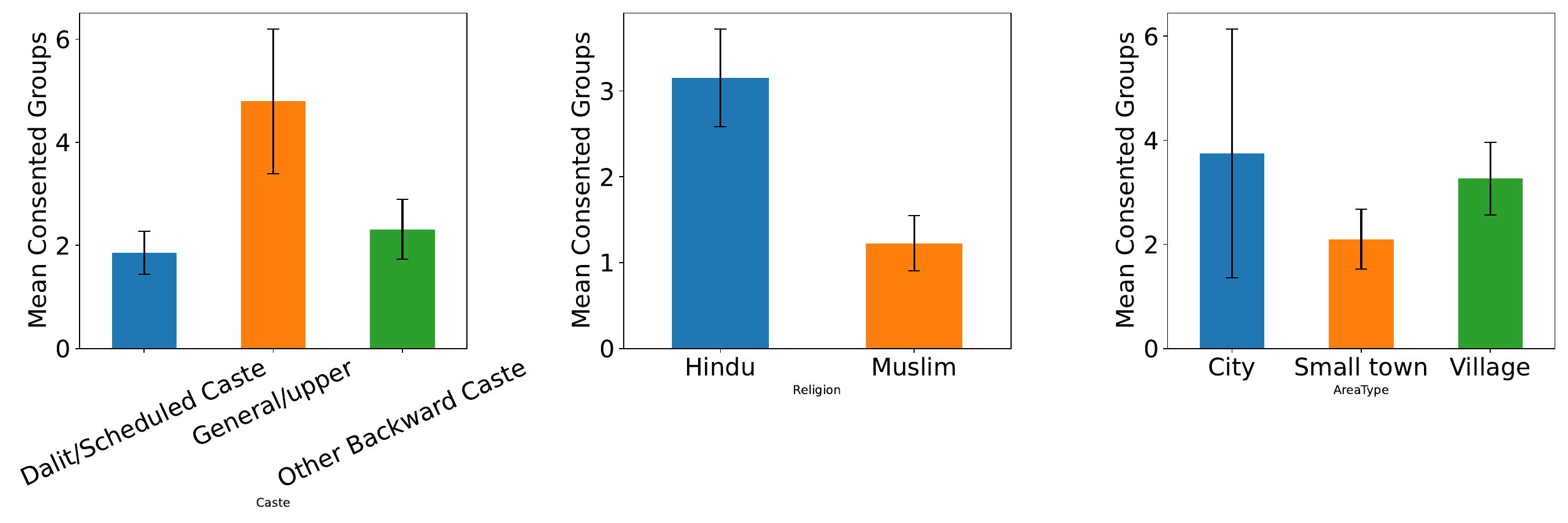}
    \caption{Demographics and consented groups (India)}
    \label{fig:demographics_consented_groups}
\end{figure}

\begin{figure}[H]
    \centering
    \includegraphics[width=1\linewidth]{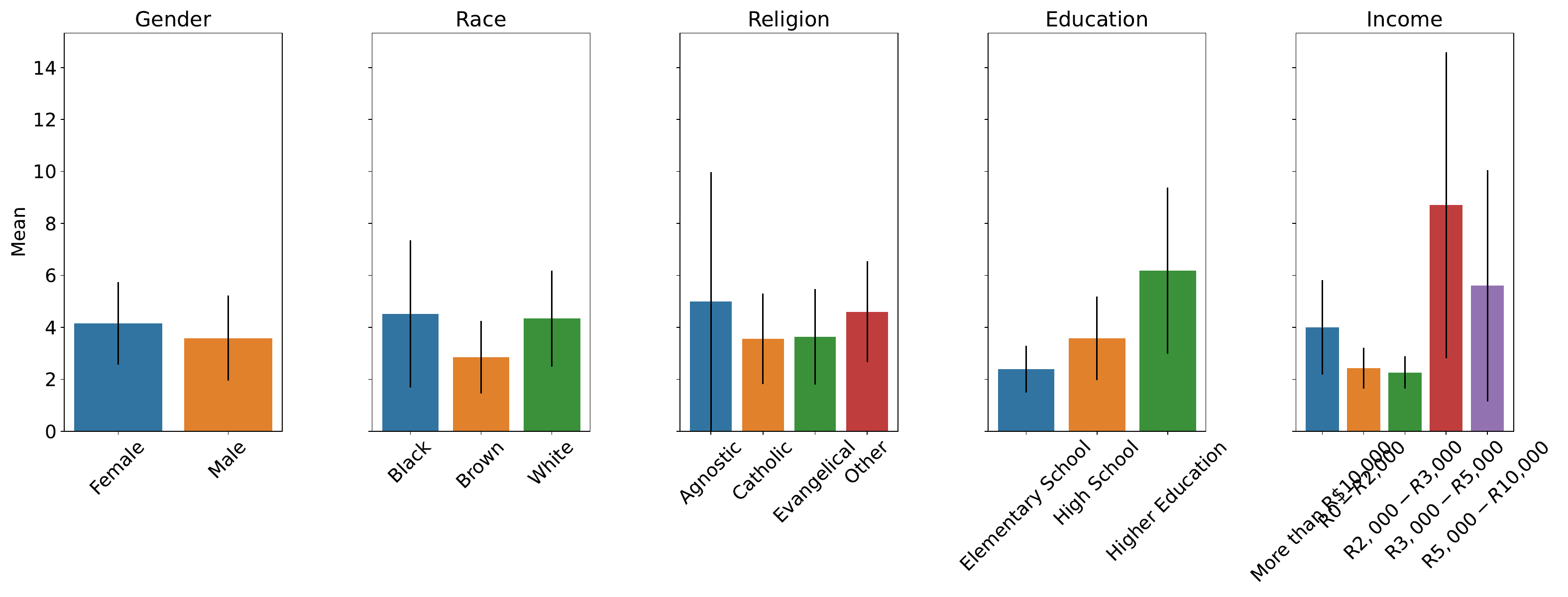}
    \caption{Brazil - mean consented groups by demographic}
    \label{fig:brazil_mean_consented_groups}
\end{figure}

\begin{figure}[H]
    \centering
    \includegraphics[width=.9\linewidth]{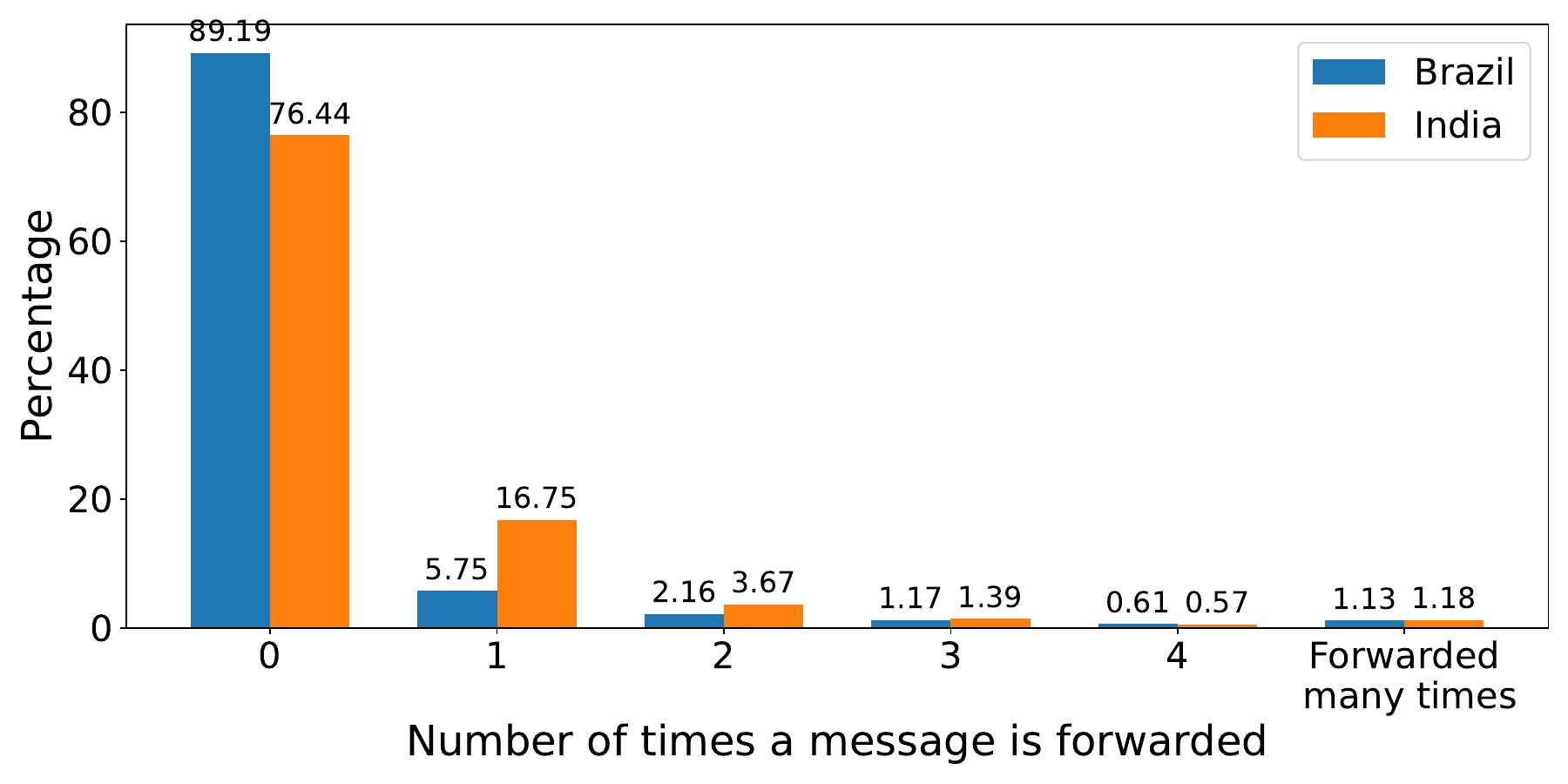}
    \caption{Forwarding score comparison. 127 indicates "forwarded many times messages". roughly 1\% in each dataset}
    \label{fig:forwarding_score_comparison}
\end{figure}

Finally, the metadata allow us to describe the type of content we extract. In Figure~\ref{fig:forwarding_score_comparison}, we first show statistics about the ratio of messages that are forwards, and the number of times they have been forwarded in a row, until they have been forwarded five times or more and are labeled as ``frequently forwarded" by WhatsApp. As shown in the Figure, the vast majority of messages are not forwards, and only around 1\% of all content is frequently forwarded content. We however note some major and interesting differences in rates of forwards across India and Brazil, with forwards accounting for a fairly large proportion of all messages received by our Indian donors.

\begin{figure}[H]
    \centering
    \includegraphics[width=.9\linewidth]{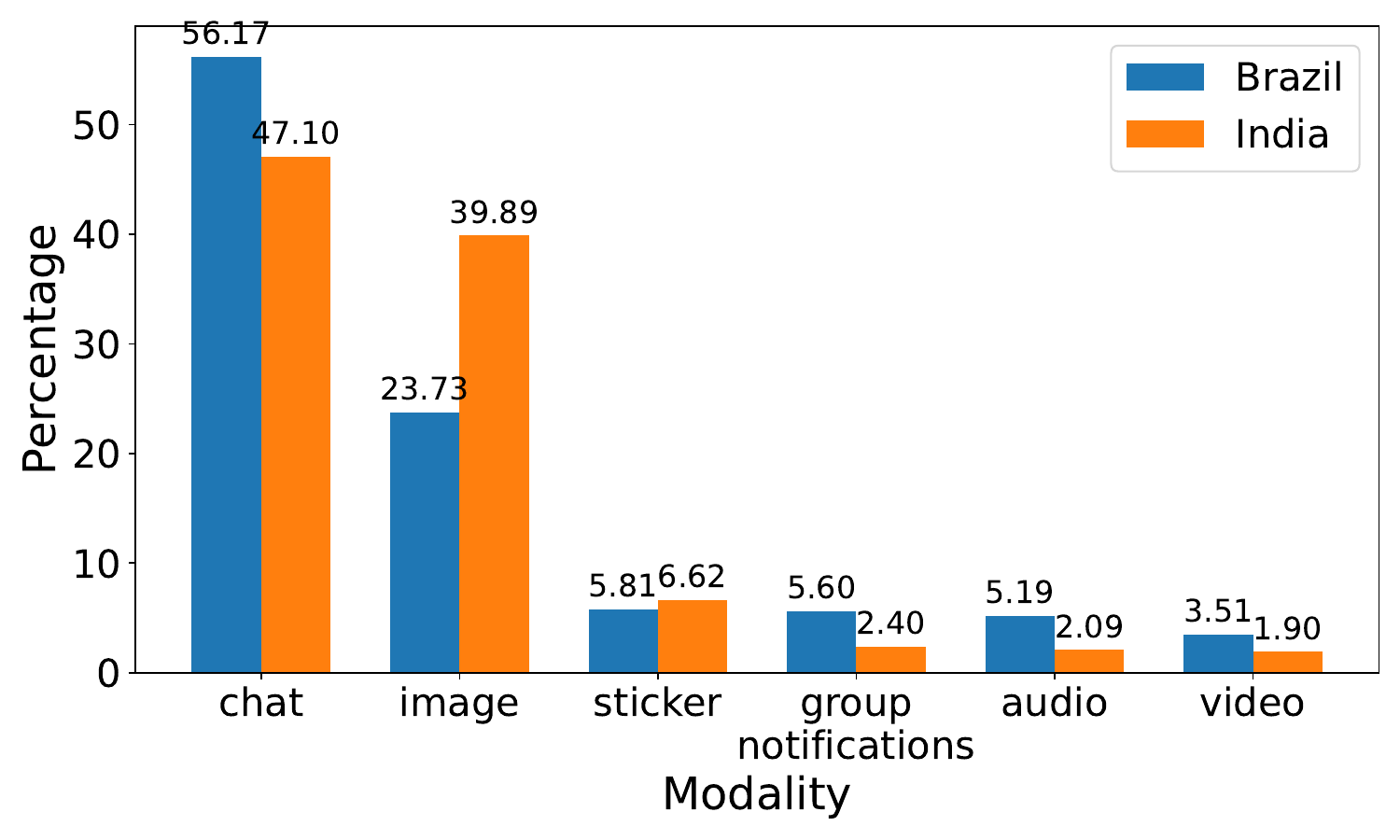}
    \caption{Comparison of the top 6 modalities}
    \label{fig:modality_comparison}
\end{figure}

Second, we can use our data to describe whether messages are chats (text messages), image, videos, or other types of content. As shown in Figure~\ref{fig:modality_comparison}, only around 50\% of all messages in both countries are chats, illustrating the multimedia nature of WhatsApp. We however note some interesting differences in the way in which Indians and Brazilians use the service. For instance, images are significantly widely used in India, with close to 40\% prevalence, Brazilians use significantly larger share of audio and video messages.

\section{Conclusion}

In sum, the strategy we detail here should provide researchers with an efficient, privacy-protecting and secure methodology to collect WhatsApp data to answer a variety of research questions. In that sense, we hope our work will help develop research on the platform and its uses around the world. 

Of course, we acknowledge that this strategy has limitations. First, we recognize that the multiple, extensive stages of anonymization we implement eventually fall short of eliminating 100\% of the possible risks of identification of the individuals involved. We however believe it comes extremely close to doing that, in practice, and note that researchers willing to undertake WhatsApp research must, in one way or another (ideally, only in a residual way), be willing to invoke that their research is in the public interest - as per GDPR's article 6 -, to deviate from otherwise very limiting guidelines and make said research practically possible. Our strategy, while it goes a long way in striking the right balance between privacy protection and our ability to do research, is not perfectly foolproof on the privacy protection front unless we invoke our right to deviate from usual privacy protection guidelines to carry research in the public interest. 

Second, we still lack sufficient data to speak to the representativeness of the data we will eventually manage to extract. Until a larger study is run, we will remain unclear as to whether the strategy will for instance function among some demographics, and the extent to which respondents will be selective in terms of the groups they choose to donate.  There is in addition little doubt that researchers focusing on populations by nature difficult to investigate (say for instance, members of a rebel army or of a vigilante group) will continue to struggle to obtain data to study the influence that WhatsApp networks may have in these processes. Our technology may not entirely change the reticence that many users may have when approached and asked to donate their smartphones’ content.   

Third and relatedly, our strategy is costly in labor, infrastructures, and resources, especially if researchers are going to provide rewards or incentives to potential donors. This implies that many researchers relying on it will not be able to collect large and/or representative datasets in the case of their choosing. 

In spite of these important limitations, we believe the technology we present in this article, and which we will keep improving over the next few years, will dramatically improve current research opportunities and practice. Our early experiments in the field in India and Brazil on our own project (ERC POLARCHATS) suggest that we will be able to obtain large datasets from a diverse, if not representative, group of users. This is, in and of itself, an improvement over the status quo, and one that should allow us to answer some important research questions and monitor the virality of problematic contents on the app. 

Further, while we acknowledge that most researchers will not be able to collect as much data as we plan to due to the rather costly nature of our strategy, we also hope it will help set the standards for how to collect WhatsApp data, regardless of the amount of data collected by specific researchers. Important discussions about consent, privacy and anonymization are at stakes and need to be balanced with the impervious need to access this data to document and analyze some pressing dangers. Even if researchers assemble datasets more limited in scope than the ones we are planning to assemble, we believe their strategy should equally go through this balancing exercise and provide clear safeguards to users. In that sense, we hope this article will push researchers to reflect on what “fair” WhatsApp data collection should look like -- a thorny issue we have tried to solve --, in addition to assisting their practical needs. 

\bibliographystyle{apsr}
\bibliography{biblio.bib}

\end{document}